\documentstyle[baasar]{article}
\begin{document}

\title{High-Impact Astronomical Observatories} 
\author{Juan P. Madrid$^1$ and F. Duccio Macchetto$^2$} 
\affil{$^1$McMaster University, Hamilton, Canada}
\affil{$^2$Space Telescope Science Institute,
3700 San Martin Dr., Baltimore, MD 21218}

\begin{abstract}

  We  derive the ranking  of the  astronomical observatories  with the
  highest impact in astronomy based on the citation analysis of papers
  published in 2006. We also  present a description of the methodology
  we use  to derive this ranking.  The current ranking is  lead by the
  Sloan Digital  Sky Survey,  followed by Swift  and the  Hubble Space
  Telescope.

\end{abstract}

\section{Introduction}

Many studies focus  on the cost of astronomical  facilities while very
little work is  done trying to evaluate the  returns of telescopes and
satellites used in astronomy (Saleh  et al. 2007). We analyze the most
cited papers published in 2006 and  cited during the last two years to
derive  a  ranking  of  the  telescopes with  the  highest  impact  in
astronomy  during that year.   Objective measurements  of productivity
and impact  are necessary  in order to  take informed  decisions about
science   policies,  scientific   directions,  funding   and  lifetime
extension  of a  given  telescope. This  particular  ranking has  been
widely used in the past to  support the work of various committees and
these results were  incorporated in their reports, a  fresh study on a
newer dataset of high-impact papers is due.

\section{Methodology}

We draw  the ranking of  the most influencial  astronomical facilities
based on  their contribution  of data  to the most  cited papers  on a
given year.  The technique  described below was implemented during the
development of a series of tools to estimate the overall impact of the
Hubble  Space Telescope  in  astronomy (Meylan  et  al.  2004).   This
particular technique of telescope  evaluation was created based on the
method set forth by Benn \& Sanchez (2001).

A detailed  account of the  method used to  derive the ranking  of the
observatories with the highest impact, and results of this exercise in
previous  years, was  given in  Madrid  \& Macchetto  (2006).  In  the
following paragraphs we will give a brief summary of the approach that
we use to generate the aforementionned ranking.

The  200 most  cited papers  published in  a given  year  constitute a
sample  large enough  to provide  a snapshot  of the  most influencial
papers published  in astronomy for  that particular year. In  fact the
200  most  cited papers  in  2006 constitute  only  0.2\%  of all  the
references  indexed by  the  ADS but  they  account for  9.5\% of  the
citations.  Moreover, as shown in  Madrid \& Macchetto (2006), the 200
most  cited papers  stand out  from the  rest of  publications  in the
distribution of citations per paper on a given year.

We obtained  the 200 most cited  papers published in  2006 through the
SAO/NASA Astrophysics Data  System (ADS).  The ADS is  the most widely
used bibliographic database in  astronomy. We went through the onerous
process of downloading each of these 200 most cited papers. Each paper
was then analyzed and we determine whether the paper was observational
or theoretical.  Theoretical papers  usually present models and do not
contain  any data  taken  with a  telescope.   On the  other hand,  an
observational  paper is  a paper  that presents  data obtained  with a
telescope or  several telescopes.   For those observational  papers we
determine which facility, or  more often, which facilities the authors
used to gather their data.

The number  of citations  of each paper  is credited to  the telescope
used to  take the  data.  The telescope  that accumulates  the largest
number of citations will thus end  up on the top of our ranking.  When
several  telescopes provide  data for  a publication  a  percentage of
participation, or  weight, is established  and a fractional  number of
citations is  credited to  each contributing facility  proportional to
their participation.  Examples of these basic arithmetics are given in
Madrid \& Macchetto (2006).

\begin{center}
\begin{tabular}{llcc}
\multicolumn{4}{c}{\sc Table 1}\\
\noalign{\smallskip}
\multicolumn{4}{c}{\sc High-Impact Observatories}\\
\\
\hline
\noalign{\smallskip}
\hline
\noalign{\smallskip}

{\bf Rank} & {\bf Facility} & {\bf Citations} & {\bf Participation}\\
\noalign{\smallskip}
\tableline
\noalign{\smallskip}

1 & SDSS     & 1892  & 14.3\% \\
2 & Swift    & 1523  & 11.5\% \\
3 & HST      & 1078  & 8.2\%  \\
4 & ESO      & 813   & 6.1\%  \\
5 & Keck     & 572   & 4.3\%  \\
6 & CFHT     & 521   & 3.9\%  \\
7 & Spitzer  & 469   & 3.5\%  \\
8 & Chandra  & 381   & 2.9\%  \\
9 & Boomerang& 376   & 2.8\%  \\
10& HESS     & 297   & 2.2\%  \\  

\noalign{\smallskip}
\hline
\end{tabular}
\bigskip
\end{center}

\section{Results}

The results of this study are summarized in Table 1 which presents the
top-ten high impact astronomical observatories. 

The  Sloan Digital Sky  Survey is  once again  the telescope  with the
highest  impact  in astronomy,  see  Madrid  \&  Macchetto (2006,  and
references  therein) for  previous rankings.   The SDSS  published its
fourth data  released in  2006.  Swift, a  telescope dedicated  to the
science of GRB, ranks second,  the findings made by this telescope has
clearly transcended  the field of GRBs  and has had a  broad impact in
astronomy.

The  Hubble  Space Telescope  ranks  third.  Several  papers from  the
Advanced  Camera for  Surveys Virgo  Cluster  Survey, as  well as  the
publication in  2006 of  the Hubble Ultra  Deep Field,  contributed to
rank HST as the third telescope with the highest impact.

Ground based optical and infrared  astronomy has an important share of
participation in the high impact papers with ESO, Keck, and the CFHT.

Spitzer and  Chandra rank  7th and  8th, it is  worth noting  that all
active  NASA Great  Observatories belong  to the  top ten  high impact
astronomical telescopes.

The High Energy Stereoscopic System  (HESS) ranks 10th. This system of
imaging atmospheric Cerenkov telescopes is located in Namibia and aims
to image very high energy  phenomena. The facilities that rank 11th to
15th and  thus do  not appear  on Table1  are WMAP,  2MASS, Gemini,
Subaru, and NOAO (CTIO+KPNO).

\acknowledgments

We  have made  extensive  use  of the  NASA  Astrophysics Data  System
Bibliographic services.


\begin{references}


\reference Benn, C. R. \& S\'anchez, S. F. 2001, PASP, 113, 385

\reference Madrid, J. P., \& F. D. Macchetto 2006, Bulletin of the
American Astronomical Society, vol 38, p. 1286

\reference Meylan, G., Madrid, J.P., \& Macchetto, D. 2004, PASP, 116,
790

\reference  Saleh,   J.,  Jordan,  N.,   Newman,  D.  J.   2007,  Acta
Astronautica, Volume 61, Issue 10, p. 889

\end{references}
\end{document}